\documentclass[aps,prb,twocolumn,groupedaddress]{revtex4-1}

\usepackage{amsmath,amssymb}
\usepackage{graphicx}
\usepackage{color, palatino, graphicx, wrapfig}

\renewcommand{\vec}[1]{{\mathbf #1}}

\begin{document}

%\preprint{}

\title{The Effect of Disorder in Superfluid Double Layer Graphene}



\author{B. Dellabetta}
\author{M.J. Gilbert}
\email[]{dellabe1@illinois.edu}
%\thanks{}
\affiliation{Department of Electrical and Computer Engineering, University of Illinois, Urbana, IL 61801}

\date{\today}

\begin{abstract}

We investigate the superfluid properties of disordered double layer graphene systems using the non-equilibrium Green's function (NEGF) formalism.  The complexity of such a structure makes it imperative to study the effects of lattice vacancies which will inevitably arise during fabrication. We present and compare room temperature performance characteristics for both ideal and disordered bilayer graphene systems in an effort to illustrate the behavior of a Bose-Einstein Condensate in the presence of lattice defects under non-equilibrium conditions.  We find that lattice vacancies spread throughout the top layer past the coherence length have a reduced effect compared to the ideal case. However, vacancies concentrated near the metal contacts within the coherence length significantly alter the interlayer superfluid transport properties.

\end{abstract}


\maketitle

\section{Introduction}

After the foundations of superconductivity were laid more than fifty years ago, the condensation of excitons, bosonic molecules comprised of one electron and one hole, in semiconductors became a topic of intense interest to physicists\cite{Keldysh:1964}. In the subsequent years, the observation of a Bose-Einstein condensate (BEC) in semiconductors has been limited by the fact that directly bound excitons have such a short lifetime\cite{Snoke:2002}. In recent years, the search for BEC in semiconducting systems has found significant experimental\cite{Tutuc:2004,Wiersma:2004,Kellogg:2004,Kellogg:2002,Wiersma:2006,Pillarisetty:2004,Plochocka:2009,Yoon:2010,TiemannC:2008,TiemannB:2008,TiemannA:2008} and theoretical progress\cite{Fertig:1989,Wen:1992,Moon:1995,Halperin:1993,Fil:2007,Park:2006,Cote:2007,Stern:2001,Balents:2001,Murthy:2008,Fogler:2001,He:1993,MacDonald:2001,Jungwirth:2001} in coupled quantum well systems in the Quantum Hall regime, where the each of the layers has a filling factor of $\nu_{layer}=\frac{1}{2}$ for a total filling factor of $\nu_{total}=1$. Despite significant efforts, however, there has yet to be any conclusive proof that this system does exhibit excitonic superfluidity. This is due to the fact that in semiconductor quantum wells, a magnetic field is required to drive the phase transition and this magnetic field creates edge states which flow along the edges of the system. Therefore, an argument may be made which states that it is just as likely that the injected quasiparticles are simply transported in the dissipationless edge states rather than in the bulk of the system as one would expect for a BEC.

Spatially separated monolayers of graphene have been predicted to exhibit excitonic superfluidity at temperatures approaching room temperature\cite{Min:2008,Zhang:2008,Gilbert:2009} before thermal fluctuations break the individual excitons and destroy the condensate. The necessary conditions required to observe excitonic superfluidity in double layer graphene are interlayer separations on the order of $1~nm$ and  an equal number of electrons and holes in the respective layers. Double layers of graphene are suited to the observation of broken symmetry states at much higher temperatures as compared to that of the semiconductor bilayers due to the fact that it is atomically two-dimensional, thereby significantly reducing the screening effects, and because it exhibits a band structure that is both linear and gapless which implies particle-hole symmetry. Experimentally speaking, this is a much more ideal system in which to observe excitonic superfluidity as double layers of graphene may nest their Fermi surfaces without the need of a magnetic field to quench the kinetic energy of the system. This eliminates the edge states which may be obstructing the clear observation of the BEC in the semiconductor layers. However, it is not at all clear that the conclusion that broken symmetry states can exist at temperatures at and above room-temperature in double layer graphene is correct. Estimates for the Kosterlitz-Thouless temperature obtained using a mean-field theory linearized version of the critical temperature ($T_c$) combined with static Thomas-Fermi screened interlayer interactions have been shown to produce a much lower transition temperature \cite{Kharitonov:2008}. This is a result of the fact that graphene has more fermionic degrees of freedom which serve to screen out the interlayer Coulomb interactions which will drive the phase transition.

%Figure 1
\begin{figure}[bp]
\includegraphics[width=3.5in]{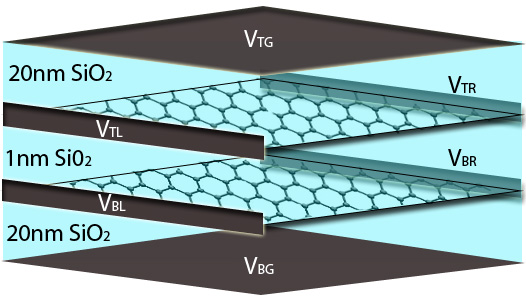}
\caption{Schematic depiction of the system we study in this work.}\label{fig:Schematic}
\end{figure}

In situations in which such discrepancies in the location of the phase boundary exist between theories, experiment will be the ultimate arbiter. While the benefits of studying such a system experimentally are clear, the issues surrounding the fabrication of such an intricate system are plentiful. As such, it is highly unlikely that the graphene layers which comprise our double layer system will be of pristine quality. Current efforts to fabricate this system have shown tantalizing clues which demonstrate that the transport properties of separated layers are indeed modified by the interplay of interlayer and intralayer Coulomb interactions\cite{Tutuc:2010}. However, the transport characteristics also do not exhibit behavior consistent with the formation of an exciton superfluid. Disorder introduced when adding the top graphene layer may well be the culprit which is currently occluding the possible phase transtion as is evidenced by the marked reduction in mobility of the top graphene layer as compared to the bottom layer. While there are numerous theoretical\cite{Ihnatsenka:2010,Adam:2009,Cresti:2009,Basu:2008,Martin:2009,Klos:2009,Yoon:2008} and experimental\cite{Gallagher:2010,Lin:2008} studies which outline the significant reductions in transport properties of single graphene layers when even small amounts of disorder are present in the system, there are currently few which consider the role of disorder in the case of double layer graphene. There are predictions which describe the role of disorder in reducing $T_{c}$ in double layer graphene\cite{Bistritzer:2009} but none which detail how the interlayer and intralayer transport in double layer graphene is modified in the presence of varying amounts of disorder present.

In this paper, we self-consistently solve the equations for electrostatics in conjunction with the non-equilibrium Green's function formalism (NEGF) and local interlayer exchange interactions to understand the evolution of separately contacted double layer graphene in the regime of excitonic superfluidity\cite{Rossi:2005,MGpseudospinPRB:2010,Su:2010}. We study this system under finite bias with varying amounts and locations of disorder. We limit the disorder in our study to be vacancies in the top graphene layer while the bottom layer is assumed to be ideal in order to better understand the situation present in the experiments. We begin in Section \ref{sec:method} by outlining the methods and approximations we employ to illustrate interlayer quantum transport near the proposed regime of excitonic superfluidity in systems where one of the layers contains disorder. In Section \ref{sec:ideal}, we present the results of our self-consistent quantum transport theory for an ideal system of nanometer sized zig-zag graphene sheets with perfect A-A registration between the top and bottom layers. We focus on the calculation and analysis of the critical tunneling currents \cite{TiemannC:2008,Tiemann:2009,Su:2010}, which denote the end of coherent interlayer transport and are marked by a significant rise in the interlayer resistance for currents beyond the critical value, in agreement with previous predictions\cite{Su:2010,MGpseudospinPRB:2010}. In Section \ref{sec:disordered}, we present the main results of our paper which are our numerical calculations for the disordered graphene system. In the calculations we present here, we treat the disorder as lattice vacancies present only in the top graphene layer. In particular, we focus on the behavior of superfluids in double layer graphene where vacancies are present in two distinct regions in our system relative to the coherence length, or the distance over which wavefunctions penetrate into the insulating superfluid state: (i) in the center of the layer past the coherence length and (ii) within the coherence length of the contacts used to inject and extract current. When disorder is only included in the regions past the coherence length in the top layer, we find that the critical current is reduced and follows a square root dependence on the top layer vacancy concentrations. However, as the disorder only makes local perturbations to the system, reducing the local exciton concentrations, we find that the the critical current is only degraded by $30\%$ of its original ideal value.  In contrast, when vacancies lie within the coherence length, the corresponding reduction in available area of conductive graphene limits interlayer conductance. This results in an imbalance in superfluid current on each side of the system, yielding a critical current less than 80\% of its ideal value and which depends linearly on the concentration of vacancies in the top layer. Furthermore, for both disorder cases we consider here, we find that no steady-state superfluid density can be found when more than 4\% vacancies are included in one monolayer.

\section{Simulation Methodology}
\label{sec:method}

In Fig. \ref{fig:Schematic}, we plot a schematic of the system of interest. In this work, we consider two zigzag graphene monolayers assumed to be perfectly aligned with one another and separated by a thin dielectric. Contacts along the edges of each of the layers inject and extract current, and top and bottom gates manipulate the quasiparticle concentrations in each of the layers. In this way, we may tune them to contain the proper quasiparticle concentrations predicted by many-body theory\cite{Min:2008,Zhang:2008,Gilbert:2009} to induce a superfluid phase transition. The top and bottom gates are separated from the graphene layers by $20$ nm of SiO$_{2}$.  The two graphene layers are separated from one another by $1$ nm SiO$_{2}$ spacer dielectric to be in the regime of superfluidity predicted by many-body calculations. We consider the oxide regions to be perfect in the sense that they do not contain any stray charges and have perfect interfaces with the graphene layers. We choose the $\hat{x}$ direction to lie along the length of the system, the $\hat{y}$ direction to lie along the width and the $\hat{z}$ along the depth. We choose each monolayer to be 30~nm long by 10~nm wide. We choose these dimensions rather than larger system sizes so that we can calculate the atomistic transport properties of an increased number of disorder distributions within a reasonable amount of time. The top and bottom gates ($V_{TG}=-V_{BG}$) are are gated to effect individual carrier concentrations of $10^{13} \text{cm}^{-2}$ in each layer, which corresponds to a Fermi energy in the top (bottom) layer of 0.4eV (-0.4eV). The gate bias conditions and the interlayer separation has been chosen so as to satisfy the conditions for room temperature pseudospin ferromagnetism\cite{Min:2008}. Here, we focus on the regime where the layer electron and hole populations place the system firmly in the dense electron-hole regime where we expect the electron-hole pairs to form a BCS-type state. This is as opposed to the dilute limit of electron-hole densities which can be described as a weakly interacting Bose system of excitons\cite{Berman:2010}. As we wish to collect a sufficient statistical distribution on the effects of vacancy distribution on the interlayer transport properties of double layer graphene, we utilize several randomized vacancy configurations for each concentration we examine in this work.

We begin the NEGF simulation \cite{SDattaQT, Datta2000253} with the atomistic tight-binding description of an individual graphene monolayer,
%Equation #1
\begin{equation}
\label{Hlayer}
H_{TL} = \sum\limits_{<i,j>} \tau\mid i\rangle \langle j\mid + V_i\mid i\rangle\langle i\mid,
\end{equation}
where lattice points $i$ and $j$ are first nearest neighbors. $\tau=-3.03eV$ is the nearest neighbor hopping energy for the $p_z$ orbital of graphene, which allows for the unique low-energy linear dispersion at the $K$ and $K'$ points in the Brillouin zone. We neglect hopping among further nearest neighbors and other orbitals, as nearest neighbor $p_z$ orbital hopping is the predominant interaction for graphene in the probed energy range. The on-site potential energy $V_i=\phi(\vec{r}_i)$ is calculated via a 3-dimensional Poisson solver. We use a phenomenological model to simulate a generic metal contact with a constant density of states\cite{10.1021/nl080255r, JGThesis}. This model captures the basic self-energy needed to appropriately simulate a metal contact without taking into account multiple orbitals or complex interface problems such as lattice mismatch or Schottky barrier height.

We may now generalize our layer Hamiltonian to the double layer Hamiltonian by coupling the top and bottom monolayers with the following Bogoliubov-de Gennes (BdG) Hamiltonian,
%Equation # 4
\begin{equation}
\label{SimpleHam}
{\cal{H}}_{BdG} = \left[ \begin{array}{cc} H_{TL} & 0 \\ 0 & H_{BL} \end{array} \right] + \sum_{\mu = x,y,z} \hat{\mu} \cdot \vec{\Delta} \otimes \sigma_\mu ,
\end{equation}
with the interlayer interactions including both single particle tunneling and the mean-field many-body contribution, $\vec{\Delta}$, coupling the two layers using a local density approximation. In Eq. (\ref{SimpleHam}), $\mu$ represents a vector that isolates each of the Cartesian components of the pairing vector, $\sigma_\mu$ represents the Pauli spin matrices in each of the three spatial directions, and $\otimes$ represents the Kronecker product.

In order to correctly account for the dynamics of the double layer graphene system, we first include the many-body interlayer interactions in the Hamiltonian. Within the Hartree-Fock mean-field approximation, we may define interlayer interactions through the expectation value of the full Hamiltonian,
%Equation # 2
\begin{equation}
\label{eq:MFApprox}
\langle \uparrow_i \mid H_{BdG} \mid \downarrow_j \rangle = U m_{exc} \delta_{i,j}.
\end{equation}
We assume that the graphene monolayers are perfectly registered, so that electrons at site $i$ in the top layer $(\langle\uparrow_i\mid)$ only bind with holes at site $j$ on the bottom layer $(\mid\downarrow_j\rangle)$ when $i=j$. $U$ is the strength of the interlayer on-site Coulomb interaction, whose selected value we address later.

$m_{exc}$ is the magnitude of the order parameter resulting from our analysis of BECs of indirectly bound excitons.  It is proportional to the off-diagonal terms in the single particle density matrix, which is represented as\cite{SDattaQT,Jung:2008}
%Equation # 3
\begin{equation}
\label{eq:DensMatrix}
\rho = \left[ \begin{array}{cc} \rho_{\uparrow\uparrow} & \rho_{\uparrow\downarrow} \\ \rho_{\downarrow\uparrow}& \rho_{\downarrow\downarrow} \end{array} \right].
\end{equation}
The on-diagonal density matrix ($\rho_{\uparrow\uparrow}, \rho_{\downarrow\downarrow}$) corresponds to the associated electron and hole densities of the top and bottom monolayers, respectively.  The order parameter $m_{exc}$ can now be defined as a function of the interlayer component of the density matrix,
%Equation # 7
\begin{equation} \label{eq:MexComponents}
\begin{aligned}
m_{exc}^x &= \rho_{\uparrow\downarrow} + \rho_{\downarrow\uparrow}=2 Re(\rho_{\uparrow\downarrow}), \\
m_{exc}^y &= -i\rho_{\uparrow\downarrow} + i\rho_{\downarrow\uparrow}=2 Im(\rho_{\uparrow\downarrow}).
\end{aligned}
\end{equation}
The density matrix is directly calculated within the NEGF formalism.  It is consequently both an input and an output of our simulation. We iterate over the above mean-field equations, in conjunction with the Poisson equation for electrostatics, to obtain a self-consistent solution with compatible particle densities and potential profile utilizing the Broyden method\cite{Broyden} to accelerate convergence.

We can now expand Eq. (\ref{SimpleHam}) to show a simplified form for our BdG Hamiltonian in which interlayer interactions are expressed in terms of their directional components\cite{MGpseudospinPRB:2010},
%Equation # 5
\begin{equation}
\label{SysHam}
{\cal{H}}_{BdG} = \left[ \begin{array}{cc} H_{TL}+\Delta_z & \Delta_x - i\Delta_y \\ \Delta_x + i\Delta_y & H_{BL}-\Delta_z \end{array} \right].
\end{equation}
The directional components of the interlayer interactions $\vec{\Delta}$ are expressed as
%Equation # 6
\begin{equation}
\label{DeltaComponents}
\begin{aligned}
\Delta_x &= (\Delta_{sas} + Um_{exc}^x) \\
\Delta_y &= Um_{exc}^y \\
\Delta_z &=  \frac{1}{2}(V_{\uparrow}\mid i \rangle \langle i \mid - V_{\downarrow} \mid i \rangle \langle i \mid).
\end{aligned}
\end{equation}
The on-diagonal term in the interlayer interactions, $\Delta_z$, is due to screening caused by the unbound carriers in each monolayer, and acts to separate the two Fermi surfaces. The value of the single particle tunneling energy, $\Delta_{sas}$, is proportional to the probability of a single electron tunneling events through the thin dielectric and recombining with a hole.  Single-particle tunneling is an adverse event, thus it is desirable to have a very thick barrier in between the two graphene layers to maximize the lifetime of the indirectly bound excitons. However, we also need strong interlayer Coulomb interactions in order to drive the superfluid phase transition which necessitates a thin barrier. In this paper, we set $\Delta_{sas}=1 \mu eV$, sufficiently small so that the lifetime of the indirectly bound exciton is long enough to observe condensation but not so small as to require an intractably large number of iterations before self-consistency is reached.

Results of previous many-body calculations show the value of the order parameter is approximately one tenth the Fermi energy \cite{Min:2008}.  Our simulations show that this value of $m_{exc}$ corresponds to an interlayer coupling strength of 2.0eV.  Although this is less than the unscreened mean-field interaction strength\cite{Jung:2008} ($U=\frac{q^2}{\epsilon d}\approx 4.6eV$), it is more plausible as it factors in damping effects due to screening\cite{Kharitonov:2008}. We set $U=2.0eV$ in all simulations for the duration of this paper. We calculate the magnitude of the order parameter using the expectation values of the density matrix in Eq. (\ref{eq:MexComponents}),
\begin{equation}
\label{MMag}
|m_{exc}|=\sqrt{(m_{exc}^x)^2+(m_{exc}^y)^2}
\end{equation}
Similarly, we may identify the phase of the order parameter dependent on the same expectation values of the density matrix\cite{MGpseudospinPRB:2010},
\begin{equation}
\label{eq:MagPhaseOP}
\phi_{exc}= \tan^{-1}\left[\frac{m_{exc}^y}{m_{exc}^x}\right].
\end{equation}
As we expect that the interlayer transport properties are similar to the case of a Josephson junction, we expect $\phi_{exc}=\pi/2$ at zero bias to maximize interlayer current\cite{Rossi:2005,Park:2006}.  The quasiparticle and condensate current densities, the artifacts of the Andreev reflection action, are proportional to the spatial phase gradient and magnitude of the order parameter\cite{Jung:2008}.

\section{Interlayer Transport in Ideal Double Layer Graphene}
\label{sec:ideal}

\begin{figure}[tp]
\includegraphics[width=3.6in]{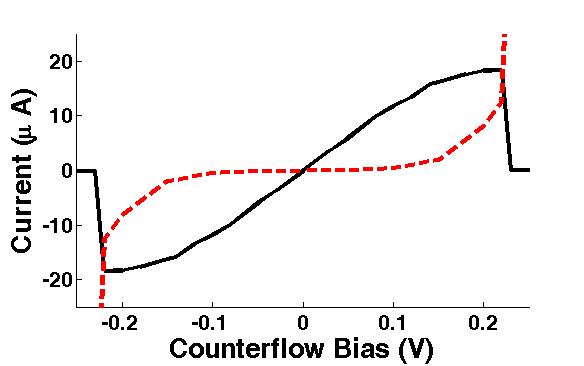}
\caption{Interlayer and intralayer current plots for the ideal system as a function of counterflow bias (solid black line and red dashed line, respectively).}\label{fig:IdealIV}
\end{figure}
\begin{figure}[tp]
\begin{center}$ \begin{array}{cc}
\includegraphics[width=3.6in]{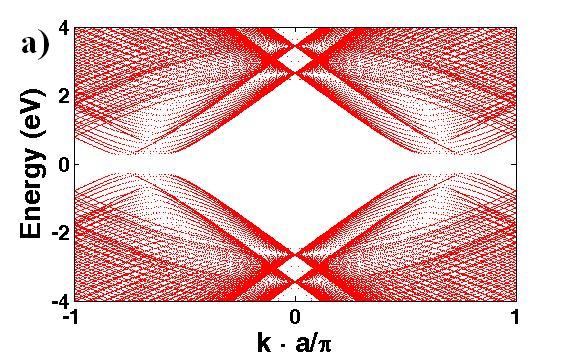} \\ \includegraphics[width=3.6in]{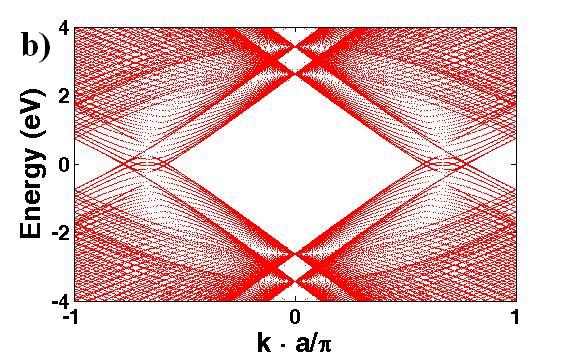} \end{array}$\end{center}
\caption{The dispersion relation for an ideal bilayer graphene system in the transport direction (a) at a counterflow bias of $V_{TL}=-V_{TR}= 0.05~V$ which is below the critical current and (b) at a counterflow bias of $V_{TL}=-V_{TR}= 0.23~V$ which is above the critical current.}\label{fig:Bstruc}
\end{figure}

It is vital to first understand the interlayer transport properties in ideal double layer graphene systems comprised of zigzag graphene nanoribbons.  After establishing ideal system properties, we can compare transport in the disordered scenarios to see how disorder impedes the inter and intralayer current flows.  To obtain a more in-depth understanding of the interlayer transport in double layer graphene systems, we draw a comparison to the Andreev reflections\cite{Andreev} that occurs at a metal-superconductor interface to explain the non-equilibrium physics of the condensate. In a superconducting system, when an electron with energy less than the superconducting band gap is injected into a superconductor, the injected electron penetrates a certain distance before producing a Cooper pair that moves across the superconductor and a retro-reflected hole of opposite spin in the normal metal.  For the exciton superfluid, an electron with energy less than the superfluid band gap injected into the system will penetrate a short distance in the top layer before driving an exciton that moves across the channel.  Put in the language of our double layer graphene system, this process corresponds to an electron being injected into the top left contact that causes an exciton to move across the system while an electron is reflected into the bottom left contact to conserve current\cite{Gilbert:2010}.

The drag-counterflow geometry ($V_{TL}=-V_{TR};V_{BL}=V_{BR}=0~V$) will cause an electron to be reflected into the bottom left contact and a hole into the bottom right contact\cite{Jung:2008}, inducing an effective current through the bottom layer.  All non-equilibrium configurations described below are in this geometry, with the left and right contacts on the top monolayer set to magnitude of the counterflow bias parameter.  The entire process results in a condensate current, due to the propagating exciton, and a quasiparticle current, caused by the injected and retro-reflected individual carriers.  The quasiparticle current is only nonzero within the coherence length $(L_c)$, the maximum length an injected particle penetrates into the superfluid gap before triggering the exciton scattering event.  The microscopic attributes of the superfluid and the dynamics of the Andreev reflection may not be readily apparent in experiments, but will have a macroscopic effect on the observable interlayer current.

In Fig. \ref{fig:IdealIV}, we present these observable interlayer and intralayer transport properties for an ideal system.  The currents are odd functions of the bias, reflecting the system's ambipolar nature. At low bias, interlayer current is linear and interlayer conductance is constant, as injected carriers see the superfluid gap and trigger the Andreev event discussed above.  Low-energy injected carriers are unable to pass through the superfluid gap, and intralayer current is negligible within this range.  States above the superfluid band gap do begin to form at $|E|\ge0.14~eV$, as seen in the superfluid band structure in Fig. \ref{fig:Bstruc}a. Interlayer transmission quickly vanishes when higher-energy states become accessible in the spectrum; transport along the monolayer dominates in these ranges. This is evidenced by the emergence of intralayer current and a drop in interlayer conductance beginning at a counterflow bias of 0.14~V.

Beyond the critical counterflow bias, 0.22$\pm$0.01~V in the ideal system, the superfluid can no longer adjust its phase to accommodate the current flow. When self-consistency is lost, only single particle tunneling contributes to interlayer tunneling.  The small magnitude of the resultant interlayer current, less than $1~pA$ for a bias of $V_{TL}=-V_{TR}= 0.25~V$, is negligible compared to the many-body contribution to the interlayer current observed when the system obtains self-consistency. Additionally, past the critical current, the transport characteristics of the bilayer becomes time-dependent\cite{Su:2010}, and the mean field equations no longer represent a valid description of the physics.  Although a transient remnant of the superfluid may yet persist beyond the critical bias, the many-body portion of the Hamiltonian is approximately zero and the time dependent oscillation in the interlayer transport relationship can not be captured in our steady-state simulations.

The phase transition is observed in Fig. \ref{fig:Bstruc}b, which shows the dispersion for the ideal system after critical current has been passed. There is no appreciable change to the band structure within each phase. When no superfluid exists, each monolayer has a vanishing band gap and linear dispersion. As expected, the Dirac points occur in the normal phase at the K' and K points in the Brillouin zone with a Fermi energy of $\pm$0.4eV.  The band gap vanishes at biases larger than this critical transition value. The closing of the band gap after the phase transition allows for more low energy states, causing a spike in intralayer current well beyond the range of interlayer transport in the superfluid phase.

The band structure in the transport direction of the system is derived\cite{Luisier:2006} from a small portion of the converged Hamiltonian by assuming $m_{exc}$ is periodic in the transport direction. This holds true only for the ideal system, where $m_{exc}$ is smooth and consistent throughout the channel with a magnitude near 10\% of the Fermi energy and a phase of $\pi/2$.  Randomly-placed impurities and vacancies, however, break translational symmetry so that the Hamiltonian is no longer periodic in the transport direction. We are thus unable to calculate the dispersion relations of disordered systems.

We find that the ideal system's critical current is $I_c=\pm 18.4 ~\mu A$, at a counterflow bias of $\pm~0.22~V$. We may compare this with the analytic approximation for the critical current in a system where coherence length is smaller than system length\cite{Su:2010}. In this case the condensate must satisfy an elliptic sine-Gordon equation,
%Equation # 8
\begin{equation}
\label{sgequation}
\lambda^{2} \vec{\nabla}^2 \phi - sin(\phi) = 0.
\end{equation}
When this equation is solved in the static case, we obtain a relatively simple expression for the critical current in our system,
%Equation 9
\begin{equation} \label{eq:CritCurrApprox}
\begin{aligned}
I_c \sim \frac{eW m_{exc}}{\hbar L_c}.
\end{aligned}
\end{equation}
Therefore, for a system of width $W=10~nm$ with a coherence length of  $L_c\approx5nm$ and order parameter magnitude of $\rho_s\approx0.04eV$, the analytic critical current is roughly $I_c^A\approx19.5\mu A$.  This is in good agreement with the critical current calculated in our simulations of an ideal bilayer. The reduction in the critical current is expected in our system as reflections off the superfluid gap lead to enhanced differences in local interlayer interaction terms which enter into the Hamiltonian self-consistently through the $\Delta_{z}$ term and serve to energetically separate the layers.\cite{MGpseudospinPRB:2010}

\begin{figure}[tp]
\includegraphics[width=3.6in]{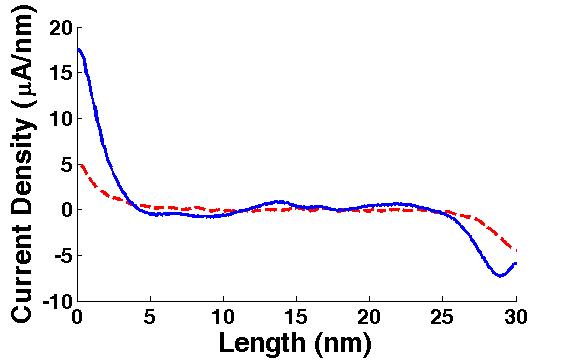}
\caption{The interlayer quasiparticle current density along the transport direction in the ideal bilayer system averaged along the width of the system for counterflow biases of 0.05V (red dashed line) and 0.20V (blue solid line).  The current density is only nonzero within the coherence length, which by inspection is approximately $5~nm$.  Quasiparticle current density is approximately odd about the center of the system at low bias, but symmetry is lost when a high contact bias limits carrier concentration on one side of the system.} \label{fig:QPcurrent}
\end{figure}

Fig. \ref{fig:QPcurrent} plots the interlayer quasiparticle current densities along the transport direction averaged along the width of the system in an ideal system with a counterflow bias of $0.05V$ and $0.20V$, just before the phase transition to a normal state. Quasiparticle current tunneling, which is part of the same process that launches the exciton through the channel, is highest near the contacts where carriers are injected, and evanescently decays into the channel, so that carriers which provoke interlayer transmission are only present a small distance into the channel. We find this distance, called the coherence length ($L_c$), to be roughly $5~nm$ by inspection of the plot.

Note in Fig. \ref{fig:QPcurrent} that interlayer quasiparticle current magnitude is nearly equivalent at low bias but becomes very asymmetric near the critical transition point. The Poisson equation in the counterflow bias configuration restricts the magnitude of the condensate order parameter because the negatively biased top right contact locally decreases the density of electrons able to pair into the condensate. Conversely, applying a bias across the hole-doped bottom layer will limit superfluid density and quasiparticle current near the left (positively biased) contact. The decreased density results in a smaller interlayer current density on one side, as seen in the case with higher counterflow bias. All interlayer currents are the currents generated by the contacts on the right side of the system, as this is the side with limited exciton density is the transport bottleneck. When the disparity in condensate currents on each side of the bilayer reaches the critical limit, steady-state supercurrents are no longer possible\cite{Jung:2008}.

\section{Interlayer Transport in Disordered Double Layer Graphene}
\label{sec:disordered}

\begin{figure}[tp]
\includegraphics[width=3.6in]{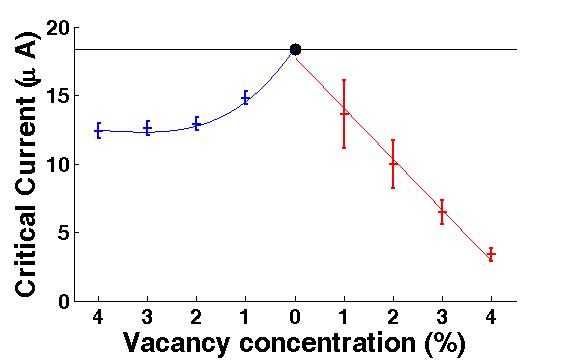}
\caption{Statistical variation of critical interlayer current for several runs at the bias just before the condensate is broken. Vacancies which are more than a distance of one coherence length away from the top layer contacts are to the left of the ideal (0\%) case, depicted by a black circle, while vacancies which are within one coherence length from the top layer contacts are to the right. The excitonic condensate is lost in both cases for top layer vacancy concentrations larger than 4\%. The solid lines represent analytic calculations using Eq. (\ref{eq:CritCurrApprox}) with modified values obtained from calculations as described in Section \ref{sec:disordered} } \label{fig:StatDist}
\end{figure}

We now seek to understand how the behavior of the interlayer transport properties change as we introduce disorder into the system. We show the main result of our analysis in Fig. \ref{fig:StatDist}, which plots the statistical variation in critical current as a function of vacancy concentration. In this figure, we show two distinct types of behavior characteristic of the two categories of locations of top layer vacancies we consider. In the middle of the figure, at a vacancy concentration of 0\%, we plot the critical current we obtain for the ideal case. To the left of the ideal case, we plot the statistical variation in the critical current for random top layer vacancy concentrations which are beyond the coherence length, $L_{c}$, from either contact on the top layer. In this case, we see that as we increase the random top layer vacancy concentrations, there is a decrease in the critical current but that the decrease is proportional to the square root of the top layer vacancy concentration. Furthermore, we also notice that the mean of the statistical distribution of critical currents associated with various top layer vacancy distributions varies only by approximately 5\% even as the concentration of vacancies is increased.

As we move away from the ideal case towards the right hand side of Fig. \ref{fig:StatDist}, we plot the statistical change of the critical current associated with within a distance of $L_{c}$ from either the left or right contact of the top layer. Here we see quite different behavior as compared to the situation where the vacancies are more than $L_{c}$ from the top layer contacts. Now the critical current has a very distinct linear decrease as we increase the vacancy concentrations. We also see that there are very large variations in the location of the critical current as we shift the locations of the top layer vacancies. In the subsequent sections, we will explore the physics of these two distinct regions and explain the observed dependence of the critical current seen in each situation.

\subsection{Layer Disorder beyond $L_{c}$}

\begin{figure}[tp]
\includegraphics[width=3.6in]{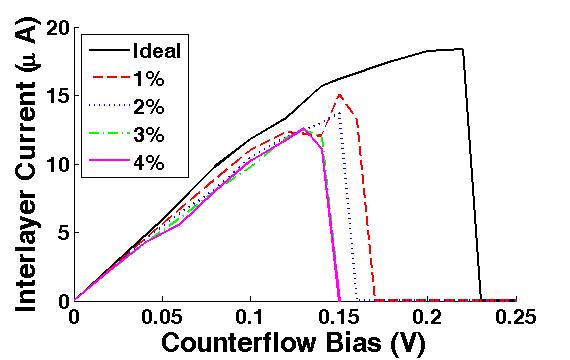}
\caption{Set of I-V curves for systems with varying concentrations of randomly placed vacancies at least 5~nm away from the top layer contacts.}\label{fig:ChIV}
\end{figure}

We model disorder as randomly placed vacancies within the top layer of our double layer system. To model a vacancy, we modify the ideal Hamiltonian by setting all hopping to a missing atom to zero. This effectively blocks any interaction with the vacancy by setting the tight-binding overlap integral of the spatial $p_z$ orbital states to zero\cite{Anantram:1998}.  We randomly remove a fixed percentage of carbon atoms from the specified region of the top monolayer, leaving the bottom monolayer unperturbed, and perform our numerical calculation.

In Fig. \ref{fig:ChIV}, we plot the interlayer current as a function of voltage bias for various concentrations of vacancies in the top layer of our double layer graphene system along with the ideal double layer case for comparison. We find that there is little reduction in the interlayer conductivity regardless of vacancy concentration or vacancy location past the coherence length at low counterflow bias. The phase transition to two normal, incoherent Fermi liquids, however, occurs at a lower bias than in the ideal case. Channel disorder decreases critical current by 20\% for 1\% vacancy concentration to 30\% for vacancy concentrations of 4\%. Beyond top layer vacancy concentrations of 4\%, we find no self-consistent solution and the interlayer transport is dominated by single particle tunneling events. Resultant interlayer currents are orders of magnitude smaller as the majority of the current injected into the system now flows across the graphene layers.

\begin{figure}[tp]
\includegraphics[width=3.6in]{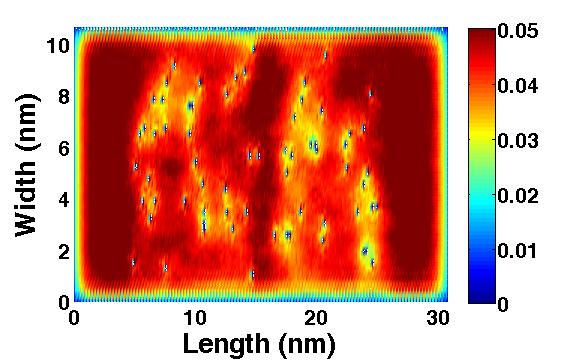}
\caption{Magnitude of the order parameter for a system with 1\% vacancies in the channel.  The condensate is very close to that of the ideal case, with a drop in magnitude only occuring locally near disorder.  As expected, $|m_{exc}|\approx0.1E_F$ away from disorder.} \label{fig:MagParamCh1D-2}
\includegraphics[width=3.6in]{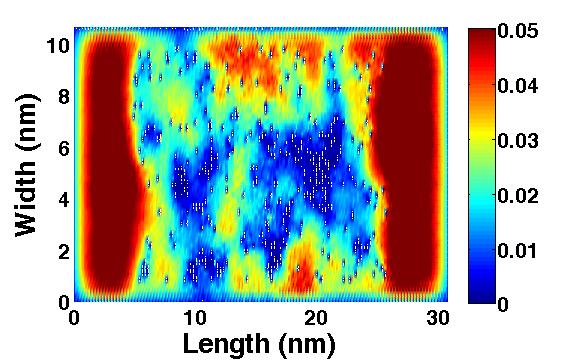}
\caption{Magnitude of the order parameter for a system with 4\% vacancies in the channel.  Magnitude clearly drops nonlinearly with the increase in disorder, but quasiparticle current is still able to form near the contacts.} \label{fig:MagParamCh4D-2}
\end{figure}

In order to better understand these reductions in interlayer current as the top layer vacancy concentration is increased, we will examine the magnitude of the order parameter, $|m_{exc}|$. In Fig. \ref{fig:MagParamCh1D-2}, we plot the magnitude of the order parameter for a random vacancy concentration of 1\% . In this situation, the order parameter magnitude does not remain constant over the entire system at 10\% $E_f$, as was the case in the ideal system. The vacancies locally destroy the condensate\cite{Kharitonov:2008} and reduces $|m_{exc}|$ at surrounding points up to a distance of $0.5~nm$, or three to four nearest neighbors from the vacancies.  However, we find no long range effect is seen when vacancies are isolated from one another by more than approximately $2~nm$. At this particular vacancy concentration, we find that $m_{exc}$ is reduced by 40\% over an appreciable area of our system as a result of these vacancies.

This situation is to be contrasted with Fig. \ref{fig:MagParamCh4D-2} where we plot $m_{exc}$ for a random top layer vacancy concentration of 4\%. In this case, significant areas clearly emerge where concentrated disorder has long range effects on superfluid density.  These areas of high vacancy concentrations in Fig. \ref{fig:MagParamCh4D-2} give rise to values for $|m_{exc}|$ that are less than 20\% the ideal value over significant areas of the system.  Superfluid magnitude remains at its ideal level near the contacts, sufficiently quarantined from the vacancies.

\begin{figure}[tp]
\includegraphics[width=3.6in]{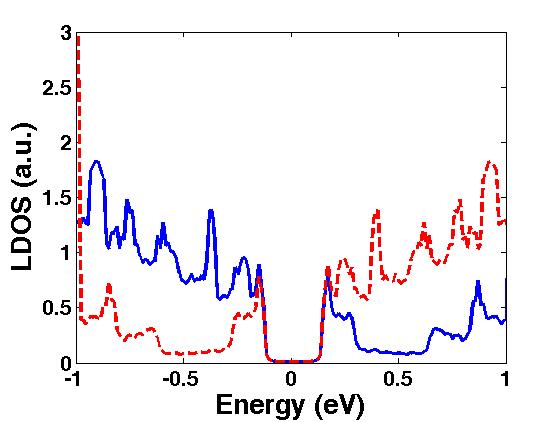}
\caption{Local density of states at low energy for an ideal system.  The solid (dashed) line represents the top (bottom) layer.  No localized states exist below a magnitude of 0.14eV for the superfluid.} \label{fig:ChLDOS0D0}
\includegraphics[width=3.6in]{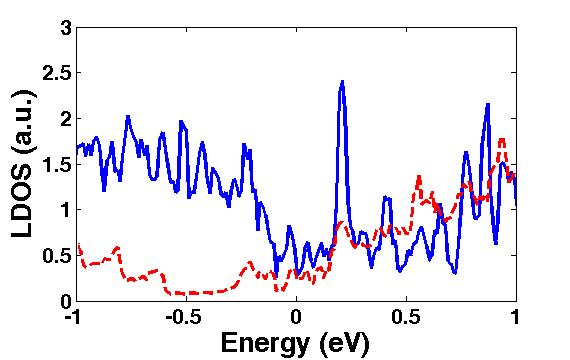}
\caption{Local density of states at low energy for a system with 4\% vacancies in the top layer.  The solid (dashed) line represents the top (bottom) layer.  Low energy states due to the vacancies introduces a weak but apparent LDOS in the ideal bottom layer as well, due to the thinness of the spacer dielectric.} \label{fig:ChLDOS4D-2}
\end{figure}

Nevertheless, the root cause of the sublinear behavior that we see in the critical current as we increase the top layer vacancy concentration is not yet resolved, from Eq. (\ref{eq:CritCurrApprox}). To explain the sublinear behavior, we examine the localized density of states (LDOS)\cite{Neophytou:2007}. Vacancies induce a LDOS similar to the case of an impurity\cite{Bacsi:2010} in graphene layers.  In Figs. \ref{fig:ChLDOS0D0} and \ref{fig:ChLDOS4D-2} we plot the LDOS for the ideal case and for the case of 4\% top layer vacancy concentration. Clearly we can see a stark contrast in the low-energy LDOS in the superfluid phase when vacancies are introduced.  The LDOS closely resembles those of monolayer graphene at higher energies\cite{Yuan:2010}, with peaks at $E=E_F\pm\tau$. The ideal top and bottom monolayers exhibit perfectly antisymmetric densities of states so that equivalent carrier concentrations arise in the oppositely gated top and bottom monolayers.  We find that no states exist in the superfluid gap in the ideal bilayer, as one would expect for a condensate in which all of the quasiparticles participate.

Disorder changes the transport properties of the system by introducing mid-gap states in the top layer seen in Fig. \ref{fig:ChLDOS4D-2}. Localized states also arise, to a lesser degree, in the bottom layer due to charge pileup induced by Coulomb attraction through the thin spacer dielectric. The biased top and bottom gates, necessary to generate the sufficient carrier concentrations, cause the undesirable occupation of the mid-gap states up to the Fermi energy. As the electrons and holes make their way across the layers, they scatter off of these localized states which changes the interlayer phase relationship between the top and bottom layers, which manifests itself as changes in the $\Delta_{y}$ term in Eq. (\ref{eq:MexComponents}). As a result, when the decreased interlayer phase relationship is input into the calculation of $m_{exc}$ in Eq. (\ref{MMag}), $m_{exc}$ drops from its ideal value of $0.041~eV$ to $0.029\pm0.002~eV$ at 4\% top layer vacancy concentration, a decrease of roughly 30\%. The result of this scattering is then that the reduced interlayer phase component then enters into Eq. (\ref{eq:CritCurrApprox}) in $m_{exc}$ it does so through the square root thereby giving rise the square root dependence that we see in on the left hand side of Fig. \ref{fig:StatDist}. When the average value of the change in the interlayer phase is added into the calculation of the critical current in Eq. (\ref{eq:CritCurrApprox}) and plotted along with our numerical results on the left hand side of Fig. \ref{fig:StatDist}, we find good agreement between the two values.

\subsection{Layer Disorder within $L_{c}$}

\begin{figure}[tp]
\includegraphics[width=3.6in]{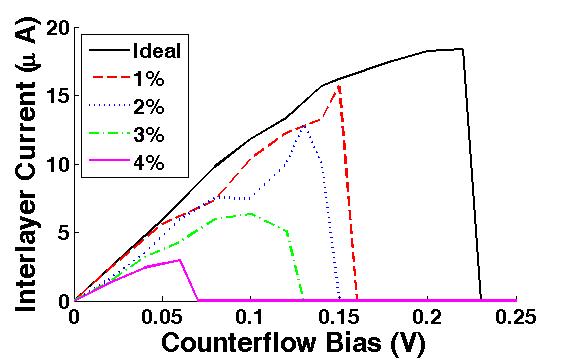}
\caption{I-V curve for systems with varying concentrations of randomly placed vacancies within 5nm of the top layer contacts.}\label{fig:SDIV}
\end{figure}

While the double layer graphene system is moderately robust to vacancies deep in the channel, this is not the case when the top layer vacancies occur within one $L_{c}$ of the contacts on the top layer. In Fig. \ref{fig:SDIV}, we plot the interlayer current as a function of the random top layer vacancy concentration where the vacancies occur within within one $L_{c}$ of the contacts on the top layer. We see that vacancies near the contacts can cause a significant drop in both conductivity and the voltage bias at which self-consistency is lost. Quasiparticle current density is now dependent on the location of disorder within the coherence length, as magnitude evanescently decays from the contact interface.  Interlayer current is thus sensitive to the specific location and amount of disorder within the coherence length.

\begin{figure}[tp]
\includegraphics[width=3.6in]{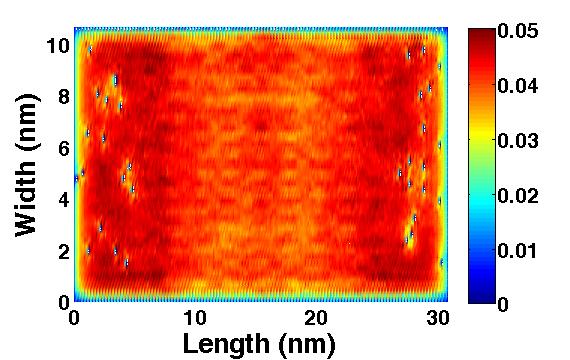}
\caption{Magnitude of the order parameter, in eV, for a system subjected to 1\% vacancies near the contacts.  A single defect does not affect the condensate appreciably, but several nearby can drop magnitude significantly.} \label{fig:MagParamSD1D-2}
\includegraphics[width=3.6in]{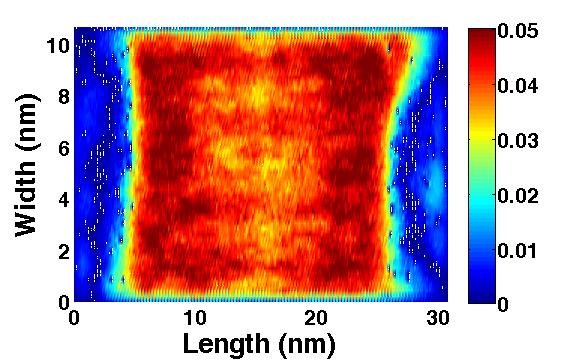}
\caption{Magnitude of the order parameter, in eV, for a system subjected to 4\% vacancies near the contacts.  The contacts are too saturated with a large amount of tunneling to occur, and interlayer conductivity suffers.} \label{fig:MagParamSD4D-2}
\end{figure}

Fig. \ref{fig:MagParamSD1D-2} shows the magnitude of the order parameter, $m_{exc}$, for a case of 1\% vacancies randomly distributed near the contacts. Quasiparticle current density drops with the presence of contact disorder, as it is proportional to the magnitude of the order parameter. Clearly, groups of vacancies appear on the left and right side of the contact that begin to show a nonlinear, long-range effect on superfluid density.  Top layer vacancies closest to the contact, where quasiparticle tunneling magnitude is largest, cause the biggest detriment to the magnitude of interlayer current. Despite the vacancies reducing the space in which quasiparticles may be injected without scattering, we still find that the condensate is able to form, and that sites within three nearest neighbors of a single defect are not appreciably affected. We find that little randomized bunching occurs in the 1\% case, as the disorder is too sparse to create significantly different scenarios. Interlayer current remains rather robust, roughly 30\% smaller than ideal.  The random nature of the placement, however, causes a high variance in interlayer current in this scenario which is discussed below.

Larger deviations from the ideal interlayer critical current are found as we move to higher top layer vacancy concentrations. We find significant amounts of clustering occur which leads to a significant reduction in the available space for quasiparticles may be injected, as seen in Fig. \ref{fig:MagParamSD4D-2}, and generates a very different transport relationship compared to the case of channel disorder. Carriers injected into the system see a significant reduction in the area in which quasiparticle tunneling can occur near the contact, significantly reducing conductivity. As we know that there is a linear dependence on the width of the system in Eq. (\ref{eq:CritCurrApprox}), this gives a very simple explanation for the physics of the linear decrease in the critical current observed in Fig. \ref{fig:StatDist}. This information allows us to conclude that graphene with vacancies within the coherence length effectively reduces the width over which quasiparticle tunneling occurs, and causes a linear decay in $I_c$ with respect to contact disorder strength until the condensate vanishes.

\begin{figure}[tp]
\includegraphics[width=3.6in]{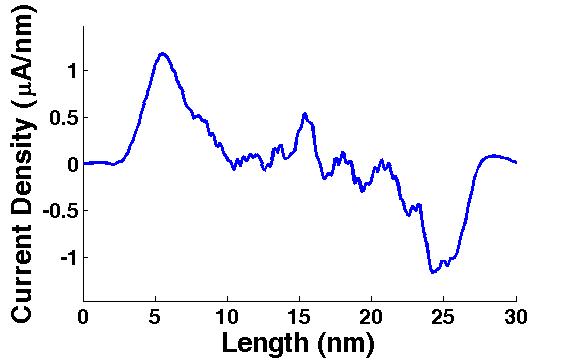}
\caption{Quasiparticle current density for a system with 4\% contact disorder with a counterflow bias of 0.04V, just before the phase transition.  No condensate exists near the contacts, so the largely degraded quasiparticle tunneling that does occur only happens beyond the disorder.} \label{fig:QPcurrentSD4D-2V4D-2}
\end{figure}

In Fig. \ref{fig:QPcurrentSD4D-2V4D-2}, we see how the average interlayer quasiparticle current density qualitatively shifts when top layer vacancies are included within the coherence length of the top layer contacts. Equivalent disorder concentrations generate disparate condensate currents on each side because vacancies are randomly configured to be more closely lumped near one contact than the other.  Whereas superfluid excitons are able to permeate disorder as long as the condensate exists, an increasing proportion of bare electrons and holes cannot penetrate the contact disorder to reach the condensate.  The critical bias at which the phase transition occurs decreases as disorder increases because of the discrepancy in condensate current from each contact.  The on-site potential is greater at vacancies near the contacts due to $V_{TR}$ and $V_{TL}$. The LDOS is thus more quickly occupied, and further accelerates the phase transition.  As a result, critical current roughly drops by an average of 30\%, 45\%, 60\%, and 80\% for vacancy concentrations of 1\%, 2\%, 3\%, and 4\%, respectively. We can see the accuracy of our conclusion on the right hand side of Fig. \ref{fig:StatDist} where we plot our data superimposed with a model of the reduction of the critical current expected when each vacancy is assumed to remove $0.5~nm$ from the effective width of the system in the effected region for a given vacancy concentration. Here we see good agreement again between our theory and numerical calculations; the model falls within the error bars of the simulations, where deviation is the result of non-linear randomized bunching of disorder.

It is clear in Fig. \ref{fig:StatDist} that a large variance in critical current exists when vacancies lie within the coherence length of the contacts in the top layer. This is due to the fact that the quasiparticle current flow, in this system, is concentrated close to the contacts in the middle of the channel, and evanescently decays into the system. This is best seen in Fig. \ref{fig:MagParamSD1D-2} where we  see that the magnitude of the order parameter is significantly reduced from the bulk value near the edge of the system both near and away from the contacts. Moreover, vacancies that are closely lumped together exacerbate the scattering to an even greater degree because vacancies in close proximity to one another have a significant non-local effect on the perturbed $m_{exc}$, as seen in Fig. \ref{fig:MagParamSD4D-2}. Also in Fig. \ref{fig:MagParamSD4D-2}, we see the variance in critical current decays at 4\% vacancies because quasiparticle tunneling is almost completely prevented at such a high concentration of vacancies.  As in the channel-disordered case, no self-consistent solution was found for vacancy concentrations greater than 4\%.

\section{Summary and Conclusions}
\label{sec:summary}

We perform self-consistent calculations to understand how vacancies in individual graphene monolayers can hinder the performance of a double layer graphene system in which a room temperature exciton condensate is predicted to form.  We find that vacancies within a coherence length of the contacts significantly obstruct performance by effectively reducing the width over which interlayer transport occurs.  For a selected value of 4\% vacancies in this region, tunneling current at a selected bias can drop by more than 80\% compared to the ideal scenario. We also find that the reduced width of the system caused by the presence of the top layer vacancies produces a linear dependence on the critical current as the vacancy concentration is increased.  Vacancies outside of the coherence length have little effect on the interlayer conductivity showing a square root dependence of the cricital current as the vacancy concentration in the top layer is increased.  Critical current degrades up to 30\% due to a phase transition at smaller bias, as $I_c$ and we find that the reduction is due to scattering of layer quasiparticles from localized mid-gap states which modifies the average interlayer phase relationship between the two layers.  Concentrations of vacancies larger than 4\% in one of the layers prevents the condensate from forming in a steady state.

\begin{acknowledgments}
The authors would like to thank J.-J. Su and A. H. MacDonald for insightful discussions as well as S. Datta, N. Neophytou and G. Liang for graphene-interface simulation methodology in the NEGF formalism.  This work is supported by ARO.
\end{acknowledgments}

\bibliography{PRBRefs}

\end{document}